# In-situ Patterned Damage-Free Etching of 3-Dimensional Structures in β-Ga$_2$O$_3$ using Triethylgallium

Nabasindhu Das[1,a], Fikadu Alema[3], William Brand[3], Abishek Katta[2], Advait Gilankar[1], Andrei Osinsky[3], Nidhin Kurian Kalarickal[1]

[1]*Department of Electrical and Computer Engineering, Arizona State University, AZ 88514 USA*

[2]*Department of Material Science & Engineering, Arizona State University, AZ 88514 USA*

[3] *Agnitron Technology Incorporation, Chanhassen, MN 55317 USA*

In this work, we report on the anisotropic etching characteristics of β-Ga$_2$O$_3$ using triethylgallium (TEGa) performed *in-situ* within an MOCVD chamber. At sufficiently high substrate temperature, TEGa can act as a strong etchant for β-Ga$_2$O$_3$ utilizing the suboxide reaction between Ga and Ga$_2$O$_3$ (4 Ga(s) + Ga$_2$O$_3$ (s) → 3Ga$_2$O (g)). We observe that due to monoclinic crystal structure of β-Ga$_2$O$_3$, TEGa etching on both (010) and (001) substrates is highly anisotropic in nature, both in terms of sidewall roughness and lateral etch rate. Smooth sidewalls are only obtained along crystal orientations that minimize sidewall surface energy. Utilizing this technique we also demonstrate deep sub-micron fins with smooth sidewalls and high aspect ratios. Furthermore, we also demonstrate the damage free nature of TEGa etching by fabricating Schottky diodes on the etched surface which display no change in net donor concentration.

Among the ultra-wide bandgap semiconductors (UWBG), β-Ga$_2$O$_3$ (4.7 eV) has garnered considerable attention due to its large breakdown field strength of 8 MV/cm, and reasonably high bulk electron mobility of ~200-250 cm$^2$/V-s.[1]-[5] These intrinsic properties translate to a high Baliga's figure of merit that is ~10X higher than SiC, making β-Ga$_2$O$_3$ a promising candidate for next generation kilo-volt class power switching. Furthermore, the availability of bulk substrates that can be grown using melt-based techniques such as Czochralski, edge defined film fed growth (EFG) and vertical Bridgman enables scalability of β-Ga$_2$O$_3$ device technology for low-cost industrial production. [6]-[11]. These attributes along with the wide doping range of $10^{15}$ cm$^{-3}$-$10^{20}$ cm$^{-3}$ have facilitated the swift advancement of β-Ga$_2$O$_3$ devices, both in vertical and lateral configurations, [10] - [15] [16] [17] [12], [13] demonstrating excellent performance.

---

[a] Author to whom correspondence should be addressed. Electronic mail: ndas11@asu.edu

Some of the promising device designs in β-Ga$_2$O$_3$ such as trench diodes, trench MOSFETs, FinFETs and heterojunction Junction barrier Schottky (JBS) diodes requires the fabrication of 3-dimensional device structures such as fins and trenches for forming channels or for field management using RESURF (reduced surface field). For three terminal devices such as trench MOSFETs and FinFETs, these 3-D structures also need to have sub-micron dimensions for effective gate control. Fabrication of such structures require etch processes that simultaneously offer controllable etch rates, smooth surface morphology, low damage and vertical sidewalls. Several investigations have taken place to explore etching of β-Ga$_2$O$_3$, including dry and wet processes. Dry etching utilizing chlorine-based reactive plasma (BCl$_3$ or Cl$_2$) has been shown to be effective in etching Ga$_2$O$_3$. However, dry etching results in substantial sub-surface damage that adversely impacts device performance, especially at high etch rates and RF powers[14], [15]. Conversely, wet etching approaches utilizing KOH, HF, and H$_3$PO$_4$ are capable of circumventing ion-induced etch damage [22]-[26]. However, wet etching is highly anisotropic in nature due to the asymmetric crystal structure of Ga$_2$O$_3$ making it difficult to realize vertical sidewalls that are necessary to form sub-micron fins and trenches. In addition to the conventional dry and wet etching techniques, metal-assisted chemical etching (MacEtch) has demonstrated efficacy in generating high aspect ratio structures [21]. However, this technique also results in slanted sidewalls and lowered Schottky barrier heights on the etched sidewalls. Therefore, an etch process that simultaneously offers controllable etch rates, is damage free and yields vertical sidewalls is currently absent in β-Ga$_2$O$_3$.

In our prior report, we showed that β-Ga$_2$O$_3$ can be controllably etched *in-situ* within an MOCVD reactor using Ga based metal organic precursors such as triethylgallium (TEGa). TEGa undergoes cracking above ~350 °C, releasing Ga adatoms on the β-Ga$_2$O$_3$ surface. Ga reduces Ga$_2$O$_3$ (Ga$^{3+}$) to Ga$_2$O (Ga$^+$) following the reaction; 4 Ga(s) + Ga$_2$O$_3$ (s) → 3Ga$_2$O (g). The volatile suboxide phase (Ga$_2$O) desorbs at high substrate temperatures, resulting in etching of the β-Ga$_2$O$_3$ layer. We demonstrated that high etch rates (>8.5 μm/hr) and smooth surface morphology (~2-3 nm roughness) can be obtained on both (001) and (010) β-Ga$_2$O$_3$ orientations [22]. In the present study, we investigate deep patterned etching of β-Ga$_2$O$_3$ using TEGa and report the anisotropic etching properties as well as the electrical characteristics of the etched surface.

Two different orientations of β-Ga$_2$O$_3$, (010) and (001) were studied in this work. Fe doped substrates were used for (010) orientation and Sn doped substrates or HVPE grown epilayers (10 μm thick, 1x10$^{16}$ cm$^{-3}$ doping) were used for the (001) orientation. Fig. 1 shows the schematic of the etch process on the sample surface. A 200 nm thick SiO$_2$ layer deposited using plasma enhanced chemical vapor deposition (PECVD) was used as the hard mask for patterned etching. The hard mask was patterned using optical lithography and the samples were cleaned using piranha to ensure removal of photoresist residue. The patterned samples were then etched in an Agnitron Agilis 100 oxide MOCVD system using TEGa as the Ga source and N$_2$ as the carrier gas. Further details regarding the etch process and its mechanism can be found in reference [27]. The samples were etched at a TEGa flow rate of 18 μmol/min, chamber pressure of 15 torr and substrate temperature of 800 °C. Post TEGa etching, the SiO$_2$ mask is removed using buffered oxide etchant (BOE) to derive the final etched structure.

We investigated the anisotropic dependence of sidewall morphology by fabricating spoke wheel structures as shown in Fig.2a and Fig.3a. The spoke wheel structures were patterned to form 3 μm wide trenches oriented along critical crystallographic in-plane directions on both the (010) and (001) substrates. We conducted deep etches of ~ 4 μm to form the spoke wheel structures and the final etch depth was measured using Dektak profilometry.

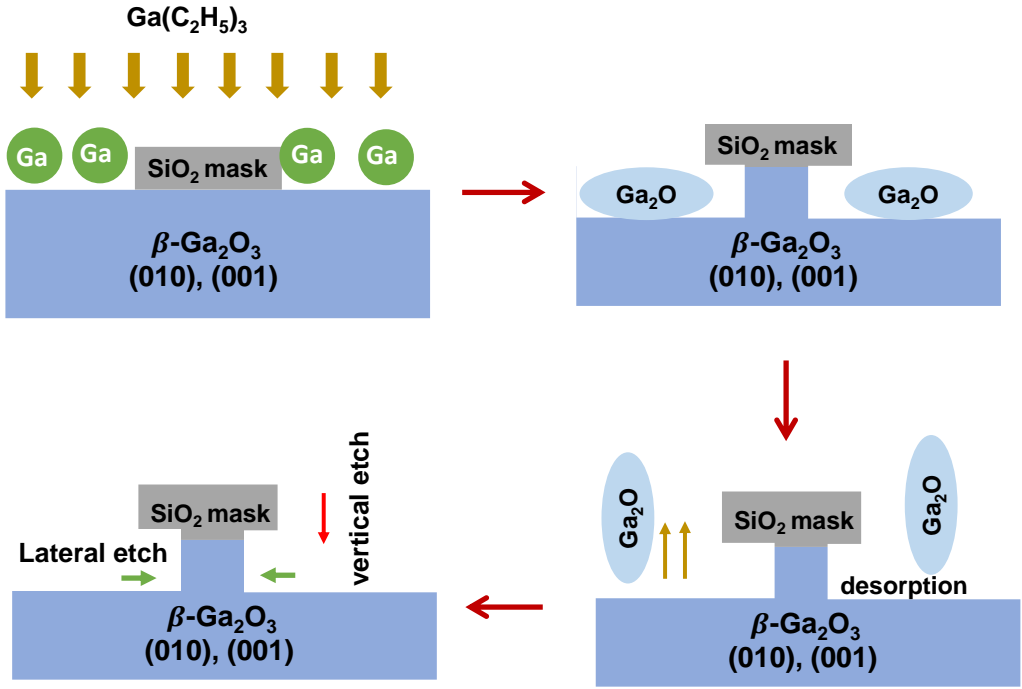

Fig 1. Schematic of MOCVD based etching of β-Ga$_2$O$_3$ using triethylgallium

We used scanning electron microscopy (SEM) with a sample tilt of ~ 40 degrees to examine the etched sidewalls. Vertical sidewalls were obtained for all in-plane trench orientations for both (010) and (001) substrates (see Fig.2 and Fig.3). Vertical sidewalls are preferred for fabricating structures such as sub-micron fins and trenches in both vertical and lateral devices since this allows uniform channel widths with high gate-to-channel aspect ratio. Vertical sidewalls were also previously obtained in MBE based Ga etching, but techniques such as MacEtch and HCl etching only show vertical (or nearly vertical) sidewalls in certain orientations [21][23]. The sidewall morphology of trenches oriented in key in-plane directions are shown in Fig.2 and Fig.3. Notably, we identify highly smooth sidewalls for trenches oriented along [001] direction (Fig 3f) on (010)- β-Ga$_2$O$_3$ substrates and [010] direction (Fig 2b) on (001)- β-Ga$_2$O$_3$ substrates respectively. On (010)-β-Ga$_2$O$_3$, [001] oriented trenches form the (100) sidewall plane, which has the lowest surface energy in β-Ga$_2$O$_3$, explaining the smooth sidewalls that are obtained [24]. Similarly, on (001)-β-Ga$_2$O$_3$, [010] oriented trenches form sidewall planes that are closest to (100) among all the in-plane sidewalls (see Fig.2b). Except the [001] direction (on (010) substrate) and [010] direction (on (001) substrate), all other trenches show significant sidewall roughening. As shown in Fig 4a, moving away from the smooth sidewall plane, we note a gradual increase in sidewall roughness, displaying a striped or stepped morphology. This anisotropy in sidewall roughness stems from the formation of (100) facets to minimize surface energy (Fig 4b-c). Consequently, moving away from the (100) sidewall plane, we observe an increase in the density of facets or steps constituting the sidewall, resulting in an enhanced stepped morphology and sidewall roughness. The faceted sidewall morphology is similar to step-terrace morphology obtained on offcut substrates as shown in Fig.4a [25].

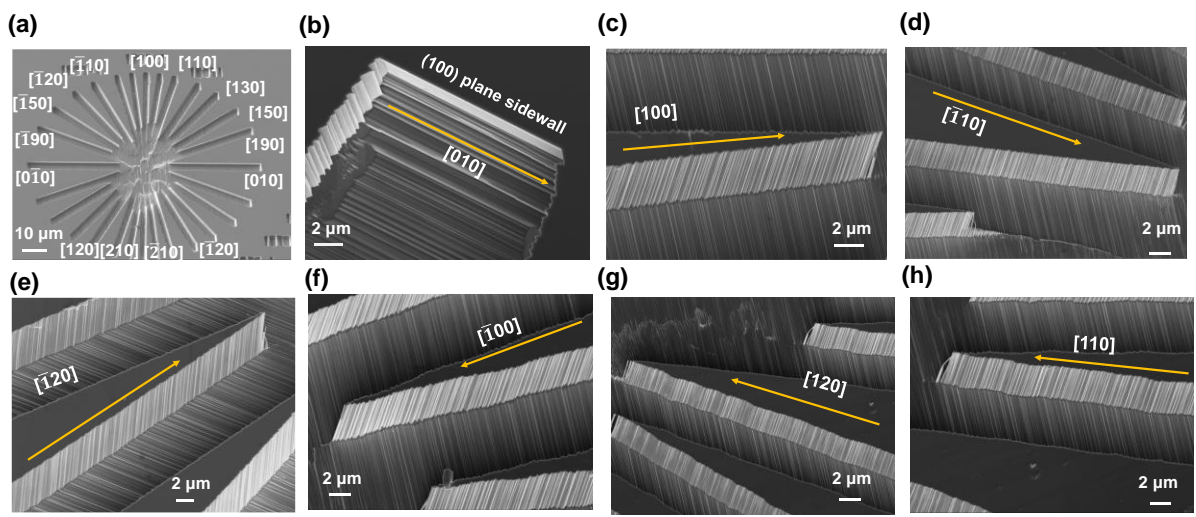

Fig 2. a) Spoke wheel structure for (001) β-Ga$_2$O$_3$ substrate b) – h) Sidewall morphology for fins in different directions

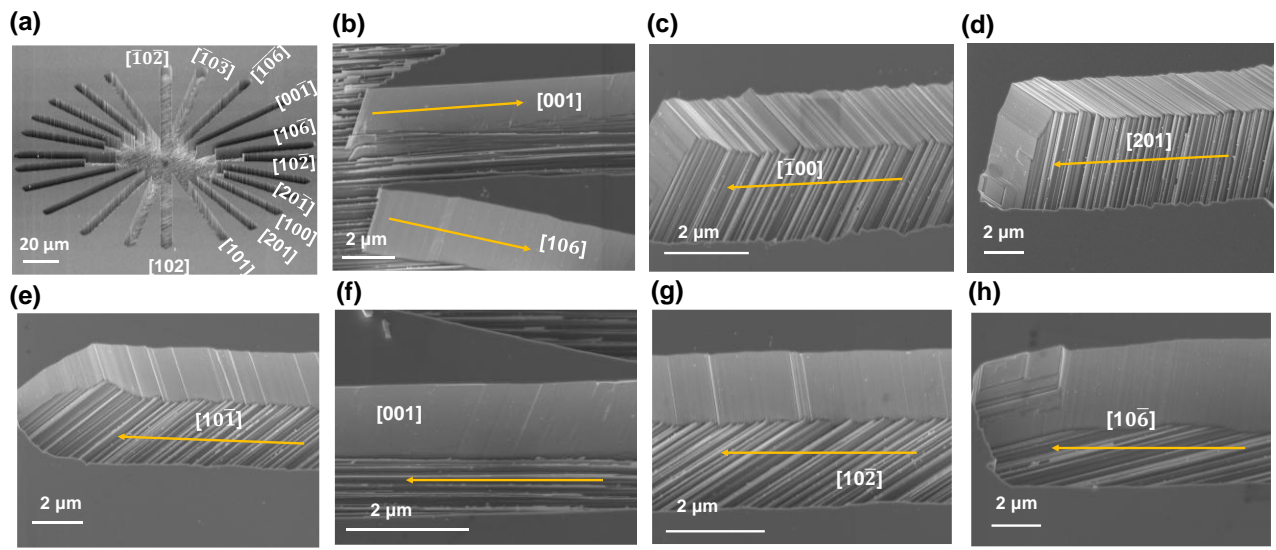

Fig 3. a) Spoke wheel structure etched on (010) β-$Ga_2O_3$ substrate. b) – h) Sidewall morphology for trenches etched in different in-plane directions on (010) β-$Ga_2O_3$ substrate.

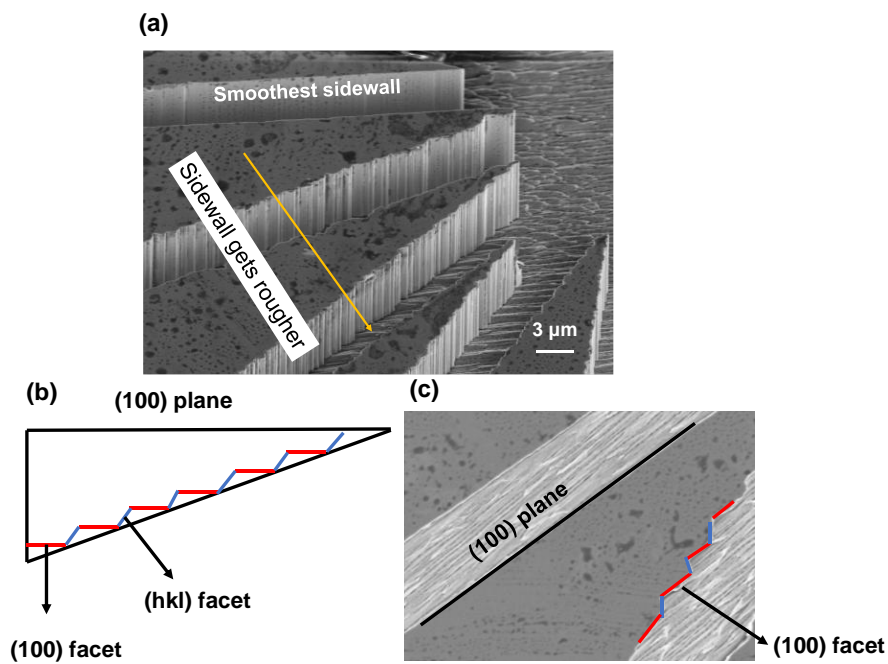

Fig 4. a) rough sidewall morphology obtained on (010) β-$Ga_2O_3$ substrate for trenches oriented at an angle to [001] direction (or (100) sidewall). b) Schematic of facets formed on the sidewall resulting in large sidewall roughness c) SEM micrograph of smooth (100) plane and rough faceted plane at an angle to the (100) plane.

In addition to vertical etching, we also observed appreciable lateral etching, which reduces (or increases) the width of features with increasing etch time. CVD is a conformal deposition process enabling deposition of Ga adatoms not only on the substrate surface, but also on the sidewalls resulting in appreciable lateral etching. Lateral etching was also previously observed in similar processes such as Ga etching in MBE and HCl etching in HVPE [23]. We observed anisotropic variation in the lateral etch rate resulting in differing final widths for trenches formed in different in-plane orientations. For most applications, it is advantageous to have low lateral etch rates since this minimizes significant variation in the final patterned structure to what was originally designed in the mask. To characterize the anisotropic dependence of lateral etch rate, we measured the initial ($w_i$) and final trench widths ($w_f$) to calculate the ratio of lateral to vertical etch rate ($r_{LV} = (w_f - w_i)/2t_v$, $t_v$=vertical etch depth). Fig. 5b and Fig. 5c shows the variation in the $r_{LV}$ ratio as a function of in-plane rotational angle. [001] trenches show the lowest $r_{LV}$ ratio of ~0.2 on the (010)-substrate, while [010] trenches show the lowest $r_{LV}$ ratio of ~0.14 on (001)-substrate. As described above, these orientations also display smooth sidewalls, making them suitable for fabricating sub-micron 3-dimensional structures. In comparison, [100] trenches showed the highest $r_{LV}$ ratio (~0.6-0.7) on both (001) and (010) oriented substrates. The significant anisotropy observed in lateral etch rates is likely due to the difference in surface energy between the different sidewall planes. The low surface energy of (100) sidewall planes make them more resistant to Ga etch when compared to sidewalls with higher surface energy.

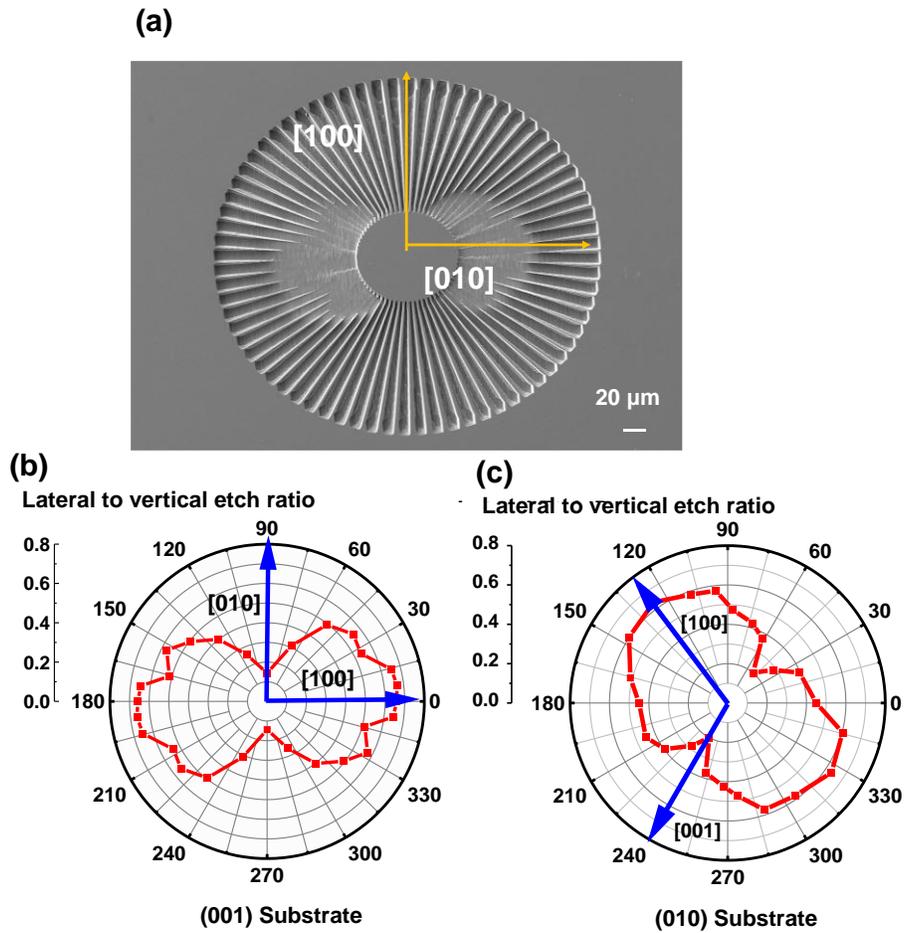

Fig 5 a) Tapered spoke wheels fabricated on (001) β-Ga$_2$O$_3$ substrate b) Ratio of lateral to vertical etch rate measured on (001) β-Ga$_2$O$_3$ substrate, c) Ratio of lateral to vertical etch rate measured on (010) β-Ga$_2$O$_3$ substrate.

In addition to using simple spoke wheel structures with constant trench width, we also fabricated tapered spoke wheel structures as shown in Fig. 5a. These structures are designed with a taper angle of 0.78 degrees, an inner trench width of 0.8 µm and outer trench width of 3 µm. The small angle taper allows amplifying a lateral etch of $l$ to a reduction in trench length by $l/\theta$. The tapered spoke wheel was also etched for the same duration as the simple spoke wheel, with a vertical etch depth of 4 µm. As shown in Fig. 5a, the etched tapered spoke wheel structure on (001)-substrate visibly reveals the in-plane variation in lateral rate, matching well with the data obtained from simple spoke wheels (Fig. 5b). Controllable lateral etching in conjunction with vertical etching using TEGa, holds promise for achieving sub-micron feature sizes without using e-beam lithography. Using this approach, we successfully fabricated sub-micrometer fins with smooth sidewalls on both (010) and (001) substrate orientations. Fig 6a depicts [010] oriented fin arrays fabricated on (001) substrate with a fin width of ~0.7 µm and fin spacing of ~3.3 µm. Fig 6b shows [001] oriented fin arrays fabricated on (010) substrate with a fin width of ~0.5 µm and fin spacing of ~3.5 µm. The [001] fin arrays had a small angle misalignment, resulting in the formation of steps which increased sidewall roughness (Fig 6b). The fins fabricated on (001) and (010) substrates show high aspect ratio of 5.7 and 8 respectively.

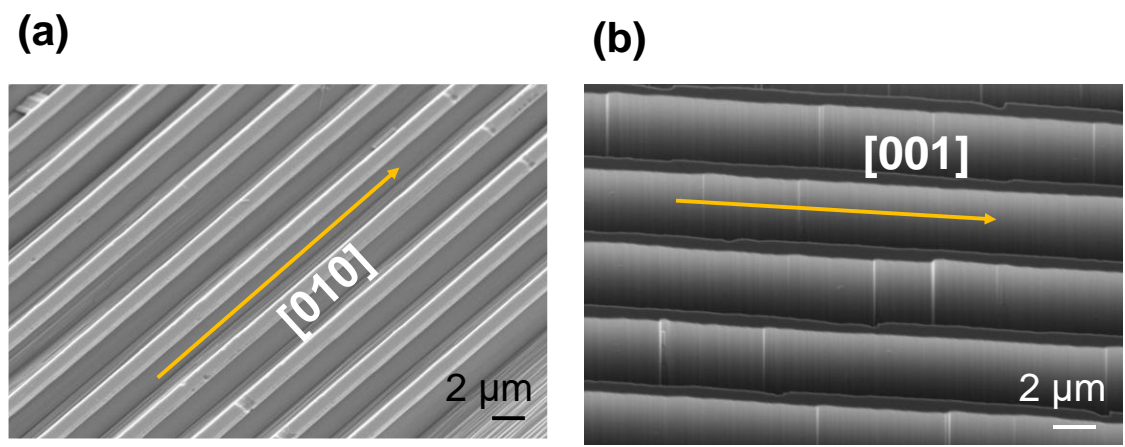

Fig 6 a)  High aspect ratio fins oriented along [010] direction on (001) β-Ga$_2$O$_3$ substrate b) High aspect ratio fins oriented along [001] direction on (001) β-Ga$_2$O$_3$ substrate.

Dopant segregation effects were observed in prior report of Ga-assisted etching implemented in MBE. This is due to the non-volatile nature of the dopants such as Si within the epilayers or substrates. Consequently, the Si from the etched epilayer segregates on the etched surface causing an increase in the surface doping concentration. To investigate potential dopant segregation and changes to the electrical properties of the etched Ga$_2$O$_3$ surface, circular Schottky contacts with a diameter of 200 µm were patterned on both etched and non-etched surface of an HVPE grown (001) epilayer ($N_D-N_A=1\times10^{16}$ cm$^{-3}$, epi thickness=10 µm) as shown in Fig 7a. The non-etched surface was masked using $SiO_2$ during the high temperature TEGa etch. A control device (Fig 7a) was also fabricated on an as-grown HVPE sample which did not see the high temperatures used for TEGa etching. Capacitance-Voltage (CV) (Fig. 5b) and current-voltage (IV) characteristics were measured and compared to study the differences between the etched, unetched and control surfaces. Figure 7d shows the comparison of dopant concentration ($N_D-N_A$) derived from C-V measurements after TEGa etching at a temperature of 800 °C, to an etch depth of 4 µm. It is observed that the net donor concentration remains unchanged after TEGa etching due to the damage free nature of the process. Fig.7 b shows the extrapolation of $1/C^2$ vs voltage measured on SBDs showing a reduction in Schottky barrier height by 0.2 eV for the unetched surface when compared to the etched surface. The unetched surface was covered with $SiO_2$ and heated to $T_{sub}$ during the in-situ TEGa etch, which likely altered the surface properties of the unetched $\beta$-$Ga_2O_3$ surface. Further investigation is required to understand the exact nature of this effect, however diffusion of Si into the $Ga_2O_3$ is not expected to be the cause, since the zero-bias depletion lengths are quite close for the etched and unetched surfaces (see Fig. 7d). Similar to C-V, a difference in Schottky barrier height of ~0.2 V is also observed between the etched and unetched surfaces in the forward I-V characteristics (Fig.7 c and Fig.7 e). The etched SBD also display a lower differential on-resistance consistent with the thinner drift layer thickness. Fig 7f shows the comparison of reverse leakage current obtained on etched, unetched and control SBDs. The reverse leakage obtained on the etched surface at -10 V ($3.7\times10^{-9}$ A/cm$^2$) is lower than that obtained on both the unetched ($7.9\times10^{-7}$ A/cm$^2$) and the control samples ($2.3\times10^{-8}$ A/cm$^2$). The lower leakage current obtained on the etched surface shows that TEGa etching is damage free and preserves a pristine $\beta$-$Ga_2O_3$ surface. The higher leakage obtained on the control sample could be potentially due to remnant surface defects from Chemical Mechanical Polishing (CMP) of the HVPE wafers [26] which is not present on the etched $\beta$-Ga$_2$O$_3$ surface. The higher leakage current obtained on the unetched surface, is consistent with the smaller Schottky barrier height obtained on this surface.

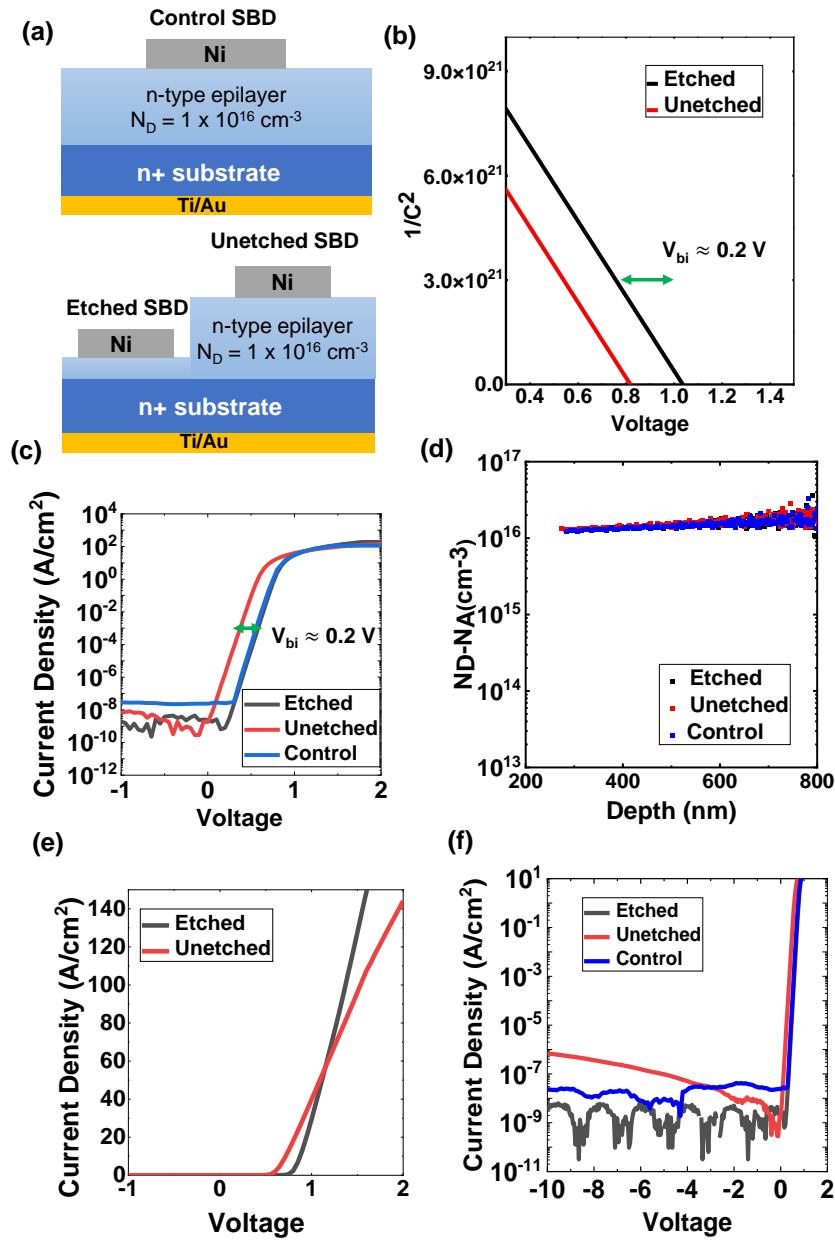

Fig 7 a) Schematic of SBDs fabricate on etched, unetched & control sample b) CV characteristics for etched and unetched devices c) Increased turn on voltage for etched SBDs d) Dopant concentration with depth for etched, unetched & control sample e) Increased Resistance for unetched SBDs f) Reverse Bias Leakage characteristics

In conclusion, we studied the anisotropic dependence of TEGa etching on both the (001) and (010) oriented β-Ga$_2$O$_3$ substrates. We find that TEGa etching forms smooth sidewalls for features oriented along [010] direction on (001)- β-Ga$_2$O$_3$ and along [001] direction on (010)- β-Ga$_2$O$_3$. Smooth sidewalls are obtained due to the formation of (100) or (100) like sidewall planes with low surface energy. Furthermore, these two in-plane orientations also display the lowest lateral etch rates with a lateral to vertical etch rate ratio of ~0.1, making them suitable for fabricating high aspect ratio 3-D structures. The damage free nature of TEGa etching was also confirmed using electrical measurements on Ni-SBDs fabricated on the etched surface which showed no change in net donor concentration ($N_D$-$N_A$). This work enhances our understanding of the in-situ TEGa etching and paves the way for the development of highly scaled vertical and lateral three-dimensional devices in β-Ga$_2$O$_3$.

**ACKNOWLEDGEMENTS**

Scanning Electron Microscopy (SEM) was performed at the John M. Cowley center for high resolution microscopy at Arizona State University. We acknowledge the use of facilities within the ASU NanoFab supported in part by NSF program NNCI-ECCS-1542160. MOCVD usage at Agnitron Technology was partially supported through a subcontract from UES through AFRL/RQKMA contract # FA8650-16-D-5408.

**Data Availability Statement**

The data that supports the findings of this study are available from the corresponding author upon reasonable request.

**BIBILIOGRAPHY**

[1] Y. Kang, K. Krishnaswamy, H. Peelaers, and C. G. Van De Walle, "Fundamental limits on the electron mobility of *β* -Ga$_2$O$_3$," *J. Phys. Condens. Matter*, vol. 29, no. 23, p. 234001, Jun. 2017, doi: 10.1088/1361-648X/aa6f66.

[2] X. Yan, I. S. Esqueda, J. Ma, J. Tice, and H. Wang, "High breakdown electric field in β-Ga2O3/graphene vertical barristor heterostructure," *Appl. Phys. Lett.*, vol. 112, no. 3, p. 032101, Jan. 2018, doi: 10.1063/1.5002138.

[3] Z. Xia *et al.*, "Metal/BaTiO3/β-Ga2O3 dielectric heterojunction diode with 5.7 MV/cm breakdown field," *Appl. Phys. Lett.*, vol. 115, no. 25, p. 252104, Dec. 2019, doi: 10.1063/1.5130669.

[4] Z. Feng, A. F. M. Anhar Uddin Bhuiyan, M. R. Karim, and H. Zhao, "MOCVD homoepitaxy of Si-doped (010) β-Ga2O3 thin films with superior transport properties," *Appl. Phys. Lett.*, vol. 114, no. 25, p. 250601, Jun. 2019, doi: 10.1063/1.5109678.

[5] J. Y. Tsao *et al.*, "Ultrawide-Bandgap Semiconductors: Research Opportunities and Challenges," *Adv. Electron. Mater.*, vol. 4, no. 1, p. 1600501, Jan. 2018, doi: 10.1002/aelm.201600501.

[6] A. Gilankar *et al.*, "Three-step field-plated β-Ga$_2$O$_3$ Schottky barrier diodes and heterojunction diodes with sub-1 V turn-on and kilovolt-class breakdown," *Appl. Phys. Express*, vol. 17, no. 4, p. 046501, Apr. 2024, doi: 10.35848/1882-0786/ad36ab.

[7] S. Paul, R. Lopez, A. T. Neal, S. Mou, and J. V. Li, "Low-temperature electrical properties and barrier inhomogeneities in ITO/β-Ga2O3 Schottky diode," *J. Vac. Sci. Technol. B*, vol. 42, no. 2, p. 024004, Mar. 2024, doi: 10.1116/6.0003401.

[8] C. Su *et al.*, "Low turn-on voltage and 2.3 kV $β$-Ga2O3 heterojunction barrier Schottky diodes with Mo anode," *Appl. Phys. Lett.*, vol. 124, no. 17, p. 173506, Apr. 2024, doi: 10.1063/5.0189890.

[9] F. Zhang *et al.*, "Mo/Au β-Ga$_2$O$_3$ Schottky Barrier Diodes With Low Turn-On Voltage and High On–Off Ratios for Low-Power Consumption Applications," *IEEE Trans. Electron Devices*, vol. 71, no. 6, pp. 3560–3564, Jun. 2024, doi: 10.1109/TED.2024.3384144.

[10] W. A. Callahan, K. Egbo, C.-W. Lee, D. Ginley, R. O'Hayre, and A. Zakutayev, "Reliable operation of Cr2O3:Mg/β-Ga2O3 p–n heterojunction diodes at 600 °C," *Appl. Phys. Lett.*, vol. 124, no. 15, p. 153504, Apr. 2024, doi: 10.1063/5.0185566.

[11] J. Y. Min, M. Labed, C. V. Prasad, J. Y. Hong, Y.-K. Jung, and Y. S. Rim, "Non-damaging growth and band alignment of p-type NiO/β-Ga$_2$O$_3$ heterojunction diodes for high power applications," *J. Mater. Chem. C*, p. 10.1039.D3TC04268E, 2024, doi: 10.1039/D3TC04268E.

[12] Y. Wei *et al.*, "Low Reverse Conduction Loss β-Ga$_2$O$_3$ Vertical FinFET With an Integrated Fin Diode," *IEEE Trans. Electron Devices*, vol. 70, no. 7, pp. 3454–3461, Jul. 2023, doi: 10.1109/TED.2023.3274499.

[13] R. H. Montgomery *et al.*, "Thermal management strategies for gallium oxide vertical trench-fin MOSFETs," *J. Appl. Phys.*, vol. 129, no. 8, p. 085301, Feb. 2021, doi: 10.1063/5.0033001.

[14] R. Khanna, K. Bevlin, D. Geerpuram, J. Yang, F. Ren, and S. Pearton, "Dry etching of Ga2O3," in *Gallium Oxide*, Elsevier, 2019, pp. 263–285. doi: 10.1016/B978-0-12-814521-0.00012-9.

[15] C. Joishi *et al.*, "Deep-Recessed $β$-Ga$_2$O$_3$ Delta-Doped Field-Effect Transistors With *In Situ* Epitaxial Passivation," *IEEE Trans. Electron Devices*, vol. 67, no. 11, pp. 4813–4819, Nov. 2020, doi: 10.1109/TED.2020.3023679.

[16] Y. Kim, J. Baek, K. H. Baik, and S. Jang, "Photochemical wet etching of (0 0 1) plane ß-phase Ga2O3, and its anisotropic etching behavior," *Appl. Surf. Sci.*, vol. 665, p. 160330, Aug. 2024, doi: 10.1016/j.apsusc.2024.160330.

[17] S. Ohira and N. Arai, "Wet chemical etching behavior of β-Ga$_2$O$_3$ single crystal," *Phys. Status Solidi C*, vol. 5, no. 9, pp. 3116–3118, Jul. 2008, doi: 10.1002/pssc.200779223.

[18] Y. Lee, N. R. Johnson, and S. M. George, "Thermal Atomic Layer Etching of Gallium Oxide Using Sequential Exposures of HF and Various Metal Precursors," *Chem. Mater.*, vol. 32, no. 14, pp. 5937–5948, Jul. 2020, doi: 10.1021/acs.chemmater.0c00131.

[19] Z. Jin *et al.*, "Wet etching in β-Ga$_2$O$_3$ bulk single crystals," *CrystEngComm*, vol. 24, no. 6, pp. 1127–1144, 2022, doi: 10.1039/D1CE01499D.

[20] H.-C. Huang, Z. Ren, C. Chan, and X. Li, "Wet etch, dry etch, and MacEtch of β-Ga2O3: A review of characteristics and mechanism," *J. Mater. Res.*, vol. 36, no. 23, pp. 4756–4770, Dec. 2021, doi: 10.1557/s43578-021-00413-0.

[21] Z. Ren *et al.*, "Temperature dependent characteristics of $\beta$-Ga2O3 FinFETs by MacEtch," *Appl. Phys. Lett.*, vol. 123, no. 4, p. 043505, Jul. 2023, doi: 10.1063/5.0159420.

[22] A. Katta, F. Alema, W. Brand, A. Gilankar, A. Osinsky, and N. K. Kalarickal, "Demonstration of MOCVD based *in situ* etching of *β*-Ga2O3 using TEGa," *J. Appl. Phys.*, vol. 135, no. 7, p. 075705, Feb. 2024, doi: 10.1063/5.0195361.

[23] T. Oshima and Y. Oshima, "Plasma-free dry etching of (001) β-Ga2O3 substrates by HCl gas," *Appl. Phys. Lett.*, vol. 122, no. 16, p. 162102, Apr. 2023, doi: 10.1063/5.0138736.

[24] S. Mu, M. Wang, H. Peelaers, and C. G. Van De Walle, "First-principles surface energies for monoclinic Ga2O3 and Al2O3 and consequences for cracking of (Al $x$ Ga1− $x$ )2O3," *APL Mater.*, vol. 8, no. 9, p. 091105, Sep. 2020, doi: 10.1063/5.0019915.

[25] Y. Zhang *et al.*, "Growth and characterization of β-Ga$_2$O$_3$ thin films grown on off-angled Al$_2$O$_3$ substrates by metal-organic chemical vapor deposition," *J. Semicond.*, vol. 43, no. 9, p. 092801, Sep. 2022, doi: 10.1088/1674-4926/43/9/092801.

[26] M. E. Liao, K. Huynh, L. Matto, D. P. Luccioni, and M. S. Goorsky, "Optimization of chemical mechanical polishing of (010) β-Ga2O3," *J. Vac. Sci. Technol. A*, vol. 41, no. 1, p. 013205, Jan. 2023, doi: 10.1116/6.0002241.